\documentclass[11pt,a4paper]{article}
\usepackage{amssymb}
\usepackage{amsmath}
\usepackage{graphics}
\usepackage{epsfig}
\usepackage{float}
\usepackage{graphicx}
\usepackage{amsfonts}
\usepackage{amsthm}
\usepackage{newlfont}
\usepackage{color}

\begin{document}

\title{Hadronic absorption cross sections of $B_{c}$}
\author{M. A. K. Lodhi$^{a}$\thanks{%
a.lodhi@ttu.edu}, Faisal Akram$^{b}$\thanks{%
faisal.chep@pu.edu.pk (corresponding author)} and Shaheen Irfan$^{b}$\thanks{%
shaheen.irfan@ciitlahore.edu.pk} \\
$^{a}$\textit{Department of Physics, MS 1051, Texas Tech University,
Lubbock\ TX 79409, USA} \\
$^{b}$\textit{Center for High Energy Physics, Punjab University, Lahore,
PAKISTAN}}
\maketitle

\begin{abstract}
\noindent The cross sections of $B_{c}$\ absorption by $\pi $ mesons are
calculated using hadronic Lagrangian based on SU(5) flavor symmetry.
Calculated cross sections are found to be in range 2 to 7 mb and 0.2 to 2 mb
for the processes $B_{c}^{+}\pi \rightarrow DB$ and $B_{c}^{+}\pi
\rightarrow D^{\ast }B^{\ast }$ respectively, when the monopole form factor
is included. These results could be useful in calculating production rate of
$B_{c}$ meson in relativistic heavy ion collisions.
\end{abstract}

PACS number(s): 13.75.Lb, 14.40.Nd, 25.75.-q

\section{Introduction}

\indent
T. Matsui and H. Satz \cite{Matsui1986} postulated that $J/\psi $ would be
dissociated due to color Debye screening in deconfined phase of hadronic
matter, called Quark-Gluon Plasma (QGP). Thus suppression of $J/\psi $ could
be regarded as a signal for the existence of QGP. NA50 experiment at CERN
\cite{NA50} has observed an anomalously large suppression of events with
moderate to large transfer energy form the Pb + Pb collision at $P_{Lab}$ = 158
GeV/c. However, this observed suppression may also occur due to absorption
by comoving hadrons. It has been argued by many authors that this phenomenon
could be significant if the absorption cross section is in the range of at
least few mb \cite%
{Cassing1997,Armesto1998,Kahana1999,Gale1999,Spieles1999,Sa1999} Extensive
work has been done to calculate these cross sections using perturbative QCD
\cite{Kharzeev1994}, QCD sum-rule approach \cite{sum-rule}, quark potential
models \cite{quark models} and hadronic Lagrangian based on flavor symmetry
\cite{Lin2001,Haglin2000,Lin2000,Liu2001}.

\noindent Bottomonium states analogous to charmonium are also subjected to
dissociation due color screening \cite{Matsui1986}, therefore their
suppression is also expected in QGP. Recently the most striking observation
from CMS (Compact Muon Solenoid experiment) is that weakly bound states of the b-quark are heavily suppressed
in Pb+Pb collisions \cite{cms}. This phenomenon is important for
understanding the properties of the QGP. Once again the knowledge of
absorption cross section is required to interpret the observed signal \cite%
{Lin2001,Vogt1997}. It has also been suggested that the production rate of
heavy mixed flavor hadrons would also be affected in the presence of QGP\cite%
{Schro2000,CUP2002}. In order to calculate production rates one require
complete knowledge of production mechanism in the presence of QGP and
absorption cross sections by comoving hadrons. In this paper we have focused
on $B_{c}$ meson. It is expected that $B_{c}$ production could be enhanced
in the presence of QGP. Due to color Debye screening, QGP contains many
unpaired $b(\overline{b})$ and $c(\overline{c})$ quarks, which upon
encounter could form $B_{c}$ and probably survive in QGP due to relatively
large binding energy \cite{Lodhi2007}. However, observed production rate
would also depend upon the absorption cross section by hadronic comovers. $%
B_{c}$ absorption cross section by nucleons has been calculated in \cite%
{Lodhi2007} using meson-baryon exchange model. This cross section is found
to have value on the order of few mb. In this paper, we have calculated $%
B_{c}$ absorption cross sections by $\pi $ mesons using hadronic Lagrangian
based on SU(5) flavor symmetry.

\noindent In Sec. II, we define hadronic Lagrangian and derive the
interaction term relevant for $B_{c}$ absorption of $\pi $ mesons. In Sec.
III, we calculate the absorption cross sections. In Sec. IV, we discuss the
numerical values of different couplings used in the calculation. In Sec. V,
we present numerical results of the cross sections with and without form factor.
Finally, some concluding remarks are made in Sec. VI.

\section{Interaction Lagrangian}

The following processes are studied in this work using SU(5) flavor
symmetric Lagrangian.%
\begin{equation}
B_{c}^{+}\pi \rightarrow DB,\text{ }B_{c}^{-}\pi \rightarrow \overline{D}%
\overline{B},\text{ }B_{c}^{+}\pi \rightarrow D^{\ast }B^{\ast },\text{ }%
B_{c}^{-}\pi \rightarrow \overline{D}^{\ast }\overline{B}^{\ast }  \label{1}
\end{equation}

\noindent First and second processes are charge conjugation of each other
and hence have same cross sections. Similarly third and fourth processes are
also charge conjugation of each other and have same cross sections.

\noindent To calculate cross sections of the above processes, we use SU(5)
flavor symmetric Lagrangian density \cite{Lin2001}. Free SU(5) Lagrangian density is given
by,%
\begin{equation}
\mathcal{L}_{0}=Tr(\partial _{\mu }P^{\dagger }\partial ^{\mu }P)-\frac{1}{2}Tr(F_{\mu
\nu }^{\dagger }F^{\mu \nu })  \label{2}
\end{equation}

\noindent Where, $F_{\mu \nu }=\partial _{\mu }V_{\nu }-\partial _{\nu
}V_{\mu } ,$ $P$ and $V_{\mu }$ denote pseudo-scalar and vector mesons
matrices as given in ref. \cite{Lin2001}.

\noindent The following minimal substitutions,%
\begin{equation}
\partial _{\mu }P\rightarrow D_{\mu }P=\partial _{\mu }P-\frac{ig}{2}[V_{\mu
},P]  \label{3}
\end{equation}%
\begin{equation}
F_{\mu \nu }\rightarrow F_{\mu \nu }-\frac{ig}{2}[V_{\mu },V_{\nu }]
\label{4}
\end{equation}%
\qquad

\noindent produce the following interaction Lagrangian desnity.%
\begin{eqnarray}
\mathcal{L} &=&\mathcal{L}_{0}+igTr(\partial ^{\mu }p[P,V_{\mu }])-\frac{g^{2}}{4}Tr([P,V_{\mu
}]^{2})  \notag \\
&&+igTr(\partial ^{\mu }V^{\nu }[V_{\mu },V_{\nu }])+\frac{g^{2}}{8}%
Tr([V_{\mu },V_{\nu }]^{2})
\end{eqnarray}%
\qquad

\noindent All mass terms, which breaks SU(5) symmetry, are added directly in
the above Lagrangian. The Lagrangian density terms relevant for $B_{c}$ absorption
by $\pi $ mesons are given by,
\begin{subequations}
\label{Lag}
\begin{eqnarray}
\mathcal{L}_{\pi DD^{\ast }} &=&ig_{\pi DD^{\ast }}D^{\ast \mu }\overrightarrow{\tau }%
\cdot (\overline{D}\partial _{\mu }\overrightarrow{\pi }-\partial _{\mu }%
\overline{D}\overrightarrow{\pi })+hc \\
\mathcal{L}_{\pi BB^{\ast }} &=&ig_{\pi BB^{\ast }}\overline{B}^{\ast \mu }%
\overrightarrow{\tau }\cdot (B\partial _{\mu }\overrightarrow{\pi }-\partial
_{\mu }B\overrightarrow{\pi })+hc \\
\mathcal{L}_{B_{c}BD^{\ast }} &=&ig_{B_{c}BD^{\ast }}D^{\ast \mu }(B_{c}^{-}\partial
_{\mu }B-\partial _{\mu }B_{c}^{-}B)+hc \\
\mathcal{L}_{B_{c}B^{\ast }D} &=&ig_{B_{c}B^{\ast }D}\overline{B}^{\ast \mu
}(B_{c}^{+}\partial _{\mu }\overline{D}-\partial _{\mu }B_{c}^{+}\overline{D}%
)+hc \\
\mathcal{L}_{\pi B_{c}D^{\ast }B^{\ast }} &=&-g_{\pi B_{c}D^{\ast }B^{\ast }}B_{c}^{+}%
\overline{B}^{\ast \mu }\overrightarrow{\tau }\cdot \overrightarrow{\pi }%
\overline{D}_{\mu }^{\ast }+hc
\end{eqnarray}
\end{subequations}
\noindent Where,

\begin{eqnarray}
D &=&\left(
\begin{array}{cc}
D^{0} & D^{+}%
\end{array}%
\right) ,\overline{D}=\left(
\begin{array}{cc}
\overline{D}^{0} & D^{-}%
\end{array}%
\right) ^{T},D_{\mu }^{\ast }=\left(
\begin{array}{cc}
D_{\mu }^{\ast 0} & D_{\mu }^{\ast +}%
\end{array}%
\right) ,  \notag \\
B &=&\left(
\begin{array}{cc}
B^{+} & B^{0}%
\end{array}%
\right) ^{T},B_{\mu }^{\ast }=\left(
\begin{array}{cc}
B_{\mu }^{\ast +} & B_{\mu }^{\ast 0}%
\end{array}%
\right) ^{T},  \notag \\
\overrightarrow{\pi } &=&\left( \pi _{1},\pi _{2},\pi _{3}\right) ,\pi ^{\pm
}=\frac{1}{\sqrt{2}}\left( \pi _{1}\mp i\pi _{2}\right)
\end{eqnarray}

\noindent Here we follow the convention of representing \ a field by the
symbol of the particle which it absorbs. The coupling constants in Eq. (6) are expressed in terms of SU(5)
universal coupling constant $g$ as following.%
\begin{equation}
g_{\pi DD^{\ast }}=g_{\pi BB^{\ast }}=\frac{g}{4},\text{ }g_{B_{c}BD^{\ast
}}=g_{B_{c}B^{\ast }D}=\frac{g}{2\sqrt{2}},\text{ }g_{\pi B_{c}D^{\ast
}B^{\ast }}=\frac{g^{2}}{4\sqrt{2}}
\end{equation}

\noindent It is also noted that SU(5) symmetry also implies the following
relation between the couplings.%
\begin{equation}
g_{\pi B_{c}D^{\ast }B^{\ast }}=2g_{\pi DD^{\ast }}g_{B_{c}B^{\ast
}D}=2g_{\pi BB^{\ast }}g_{B_{c}BD^{\ast }}
\end{equation}

\section{$B_{c}$ absorption cross section}

\begin{figure}[!h]
\begin{center}
\includegraphics[angle=0,width=0.50\textwidth]{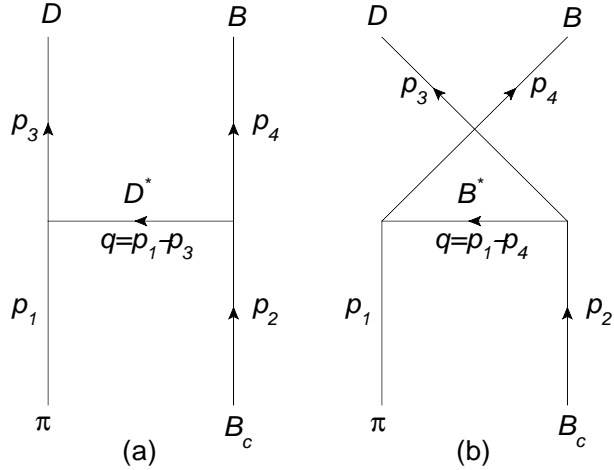}
\end{center}
\caption{Feynman Diagrams for $B_{c}$ absorption process $B_{c}^{+}\protect%
\pi \rightarrow DB$.}
\label{fig1}
\end{figure}

Feynman diagrams of the process $B_{c}^{+}\pi \rightarrow DB$ are shown in
Fig. \ref{fig1}

\begin{figure}[!h]
\begin{center}
\includegraphics[angle=0,width=0.70\textwidth]{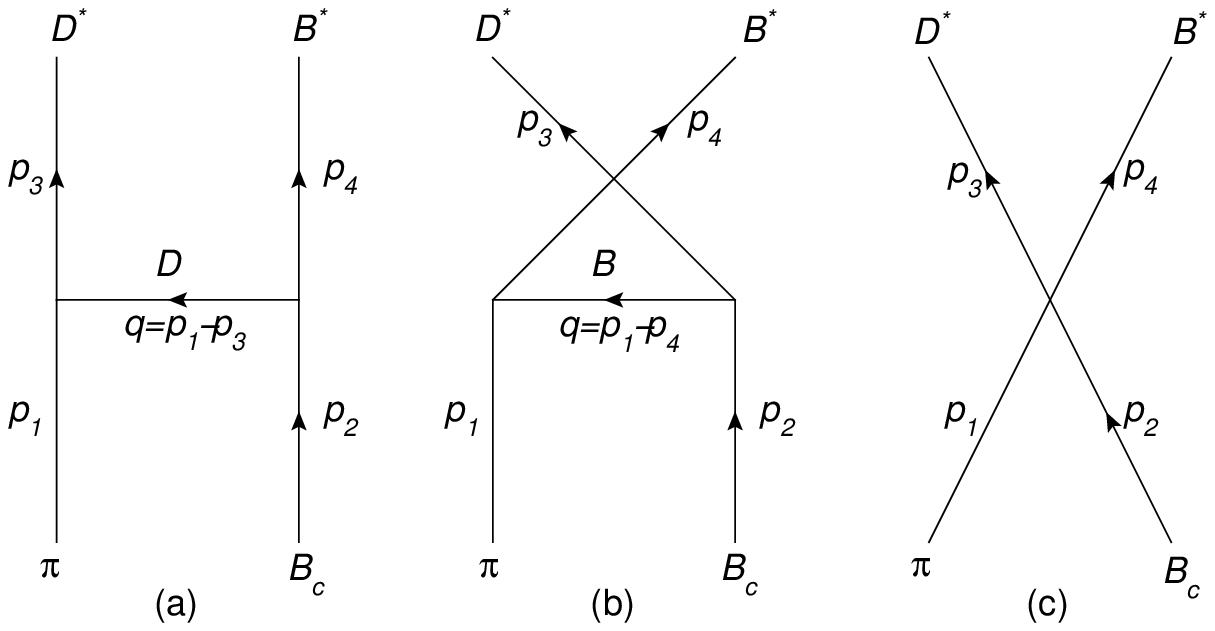}
\end{center}
\caption{Feynman Diagrams for $B_{c}$ absorption process $B_{c}^{+}\protect%
\pi \rightarrow D^{\ast }B^{\ast }.$}
\label{fig2}
\end{figure}

\bigskip

\noindent Scattering amplitudes of these diagrams are given by,
\begin{subequations}
\begin{eqnarray}
M_{1a} &=&g_{\pi DD^{\ast }}g_{B_{c}BD^{\ast }}(p_{1}+p_{3})_{\mu }\frac{-i}{%
t-m_{D^{\ast }}^{2}}\left( g^{\mu \nu }-\frac{(p_{1}-p_{3})^{\mu
}(p_{1}-p_{3})^{\nu }}{m_{D^{\ast }}^{2}}\right) (-p_{4}-p_{2})_{\nu } \\
M_{1b} &=&g_{\pi BB^{\ast }}g_{B_{c}B^{\ast }D}(p_{1}+p_{4})_{\mu }\frac{-i}{%
u-m_{B^{\ast }}^{2}}\left( g^{\mu \nu }-\frac{(p_{1}-p_{4})^{\mu
}(p_{1}-p_{4})^{\nu }}{m_{B^{\ast }}^{2}}\right) (-p_{3}-p_{2})_{\nu }
\end{eqnarray}

\noindent Total amplitude is given by,
\end{subequations}
\begin{equation}
M_{1}=M_{1a}+M_{1b} \label{11}
\end{equation}

\noindent Feynman diagrams of the process $B_{c}^{+}\pi \rightarrow D^{\ast
}B^{\ast } $ are shown in Fig. \ref{fig2}

\bigskip

\noindent Scattering amplitudes of these diagrams are given by,
\begin{subequations}
\begin{eqnarray}
M_{2a} &=&-g_{\pi DD^{\ast }}g_{B_{c}B^{\ast }D}(2p_{1}-p_{3})_{\mu }\frac{i%
}{t-m_{D}^{2}}(p_{2}-p_{1}+p_{3})_{\nu }\varepsilon _{r}^{\mu
}(p_{3})\varepsilon _{s}^{\nu }(p_{4}) \\
M_{2b} &=&-g_{\pi BB^{\ast }}g_{B_{c}BD^{\ast }}(2p_{1}-p_{4})_{\mu }\frac{i%
}{u-m_{B}^{2}}(p_{2}-p_{1}+p_{4})_{\nu }\varepsilon _{r}^{\mu
}(p_{3})\varepsilon _{s}^{\nu }(p_{4}) \\
M_{2c} &=&-ig_{\pi B_{c}B^{\ast }D^{\ast }}g_{\mu \nu }\varepsilon _{r}^{\mu
}(p_{3})\varepsilon _{s}^{\nu }(p_{4})
\end{eqnarray}

\noindent And total amplitude is given by,
\end{subequations}
\begin{equation}
M_{2}=M_{2a}+M_{2b}+M_{2c} \label{13}
\end{equation}

\noindent Using the total amplitudes given in Eqs. \ref{11} and \ref{13}, we calculate unpolarized but not the isospin averaged cross sections. The isospin factor in this case is simply 2 for the both processes.

\section{Numerical values of input parameters}

Numerical values of all the masses are taken from Particle Data Group \cite%
{PDG}. The coupling constant $g_{\pi DD^{\ast }}=4.4$, is determined
from $D^{\ast }$ decay width \cite{Col1994,Bel1995}. The coupling $g_{\pi
BB^{\ast }}$ can be fixed by two methods. Heavy quark symmetries \cite%
{Bel1995,Wise1992,Grin1993} imply that $g_{\pi BB^{\ast }}\approx g_{\pi
DD^{\ast }}\frac{m_{B}}{m_{D}}=12.4$ and from light-cone QCD sum rule \cite%
{Bel1995}, we obtain $g_{\pi BB^{\ast }}=10.3$ . In this paper, we use the
value obtained from the former method.

\noindent The values of the couplings $g_{B_{c}BD^{\ast }}$ and $%
g_{B_{c}B^{\ast }D}$ are fixed by using $g_{\Upsilon BB}=13.3$, which is
obtained using vector meson dominance (VMD) model in ref. \cite{Lin2001} and SU(5)
symmetry result $g_{B_{c}BD^{\ast }}=g_{B_{c}B^{\ast }D}=\frac{2}{\sqrt{5}}%
g_{\Upsilon BB}$ \cite{Lodhi2007}. In this way we obtain $g_{B_{c}BD^{\ast
}}=g_{B_{c}B^{\ast }D}=11.9$.

\noindent There is no empirically fitted value available for the four-point coupling $%
g_{\pi B_{c}B^{\ast }D^{\ast }}$, thus we use SU(5) symmetry, which implies $%
g_{\pi B_{c}D^{\ast }B^{\ast }}=2g_{\pi DD^{\ast }}g_{B_{c}B^{\ast
}D}=2g_{\pi BB^{\ast }}g_{B_{c}BD^{\ast }}$. These two identities give two
values of 105 and 295, whereas their mean values in 200. The values of coupling
constants used in this paper and methods for obtaining them are summarized in Table \ref{table1}.

\begin{table}[!h]
\begin{center}
\begin{tabular}{|c|c|c|}
\hline
Coupling constant & Value & Method of Derivation \\ \hline
$g_{\pi DD^*}$ & 4.4 & $D^*$ decay width \\
$g_{\pi BB^*}$ & 12.4 & Heavy quark symmetries \\
$g_{B_cBD^*}$ and $g_{B_cB^*D}$ & 11.9 & VMD, SU(5) symmetry \\
$g_{\pi B_cB^*D^*}$ & 105 to 295 & SU(5) symmetry \\ \hline
\end{tabular}%
\end{center}
\caption{Values of coupling constants used in this paper}
\label{table1}
\end{table}

\bigskip

\section{Results and Discussion}

Fig. 3 shows the $B_{c}$ absorption cross sections of the process $B_{c}^{+}\protect%
\pi \rightarrow DB$ as a function of total center of mass (c.m) energy $\sqrt{s}$.
Solid and dashed curves in this figure represent cross sections without and with form
factors. Form factors are included to account the finite size of interacting
hadrons. We use following monopole form factor at three point vertices.%
\begin{equation}
f_{3}=\frac{\Lambda ^{2}}{\Lambda ^{2}+\overline{q}^{2}}
\end{equation}

\noindent Where, $\Lambda $ is cutoff parameter and $\overline{q}^{2}$ is
squared three momentum transfer in c.m frame. At four point vertex, we use
the following form factor.%
\begin{equation}
f_{4}=\left( \frac{\Lambda ^{2}}{\Lambda ^{2}+\overline{q}^{2}}\right) ^{2}
\end{equation}

\noindent Where, $\overline{q}^{2}=\frac{1}{2}\left[ (\overline{p}_{1}-%
\overline{p}_{3})^{2}+(\overline{p}_{1}-\overline{p}_{4})^{2}\right] _{c.m}$

\begin{figure}[!h]
\begin{center}
\includegraphics[angle=0,width=0.70\textwidth]{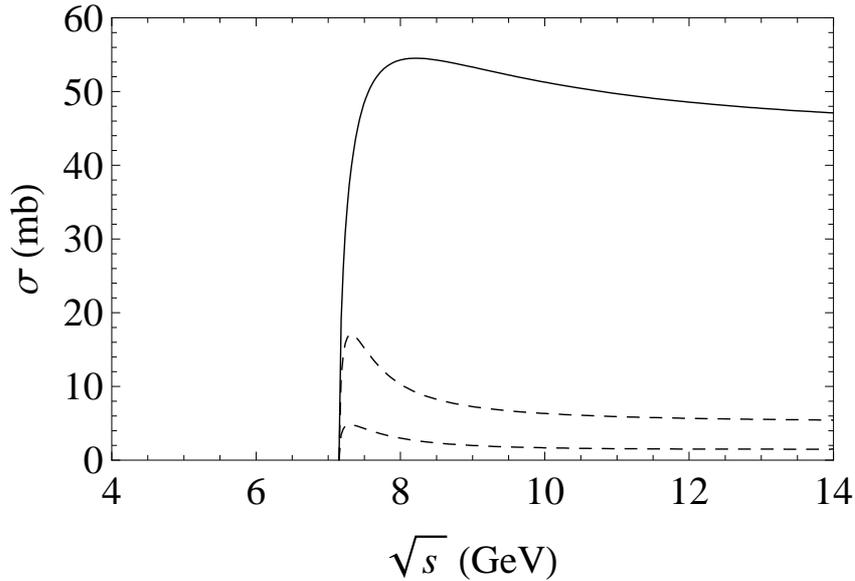}
\end{center}
\caption{$B_{c}$ absorption cross sections for the process $B_{c}^{+}\protect%
\pi \rightarrow DB$. Solid and dashed curves represent cross sections
without and with form factor respectively. Lower and upper dashed curves are
with cutoff parameter $\Lambda =1$ and $2$ GeV respectively. Threshold
energy is 7.15 GeV.}
\label{fig3}
\end{figure}

\bigskip

\noindent In general, the value of cutoff parameter used in the form factor could
have different values at different vertices. There is no direct
way to calculate the values of these parameters. In some cases cutoff
parameters can be fixed empirically by studying hadronic scattering data in
meson or baryon exchange models. Such empirical fits put the cutoff
parameters on the scale of 1 to 2 GeV for the vertices connecting light
hadrons ($\pi$, $K$, $\rho$, $N$ etc) \cite{machleid1987}. However, due to limited
information about the scattering data of charmed and bottom hadrons, no
empirical values of the related cutoff parameters are known. In this case we
can estimate cutoff parameters by relating them with inverse (rms) size of
hadrons. Cutoff parameter for meson-meson vertex is determined by the ratio of
size of nucleon to pseudoscalar meson in ref. \cite{yasui2009}.

\begin{equation}
\Lambda _{D}=\frac{r_{N}}{r_{D}}\Lambda _{N},\text{ \ \ \ \ \ \ \ }\Lambda
_{B}=\frac{r_{N}}{r_{B}}\Lambda _{N}
\end{equation}

\noindent The values of the ratios $r_{N}/r_{D}=1.35$ and $r_{N}/r_{B}=1.29$
are determined by the quark potential model for $D$ and $B$ mesons
respectively \cite{yasui2009}. Cutoff parameter $\Lambda _{N}$ for nucleon-meson vertex can
be determined from empirical data of nucleon-nucleon system. In ref. \cite{yasui2009} $%
\Lambda _{N}=0.94$ GeV, is fixed from the empirical value of the binding
energy of deuterium. Where as, nucleon-nucleon scattering data gives $%
\Lambda _{\pi NN}=1.3$ GeV and $\Lambda _{\rho NN}=1.4$ GeV \cite{holinde1989}. A variation
of 0.9 to 1.4 GeV in $\Lambda _{N}$ produces variation of 1.2 to 1.8 GeV in $%
\Lambda _{D}$ and $\Lambda _{B}$. Based on these results we take all the
cutoff parameters same for simplicity and vary them on the scale 1 to 2 GeV
to study the uncertainties in cross sections due to cutoff parameter.

\noindent Fig. 3 shows that for $B_{c}^{+}\pi \rightarrow DB$ process the cross section roughly varies
from 2 to 7 mb, when the cutoff parameter is between 1 to 2 GeV. Suppression
due to form factor at cutoff $\Lambda =1$ and 2 GeV is roughly by factor 11
and 3 respectively.

\begin{figure}[!h]
\begin{center}
\includegraphics[angle=0,width=0.45\textwidth]{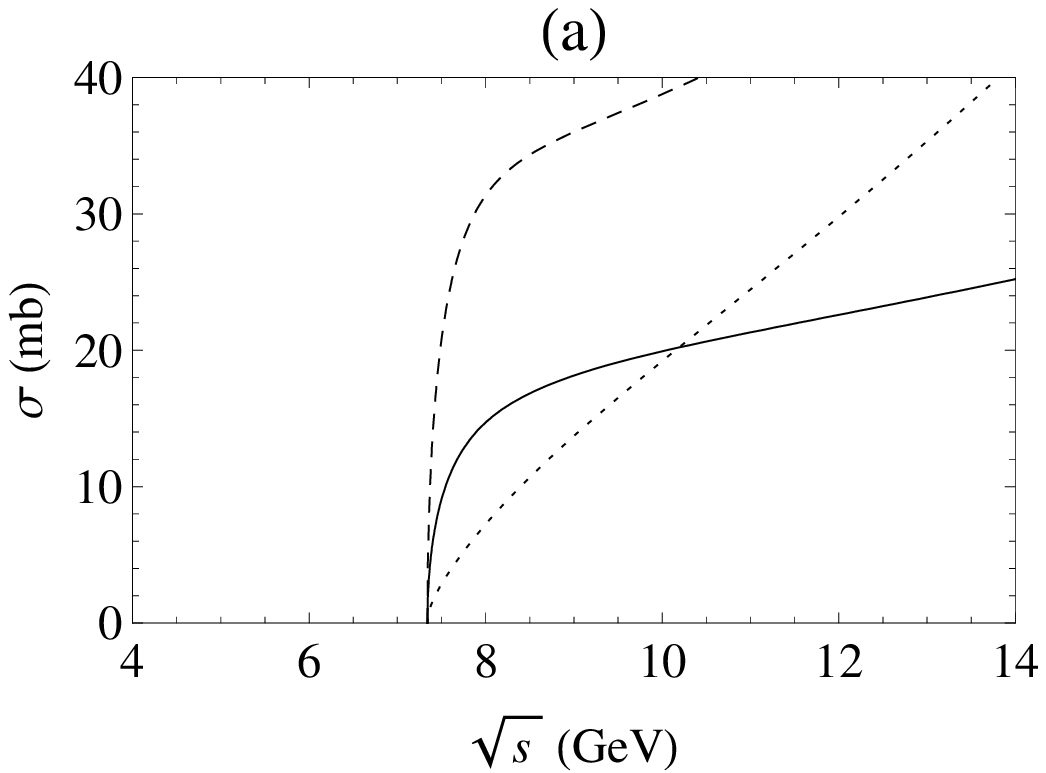} \label{fig4a} %
\includegraphics[angle=0,width=0.45\textwidth]{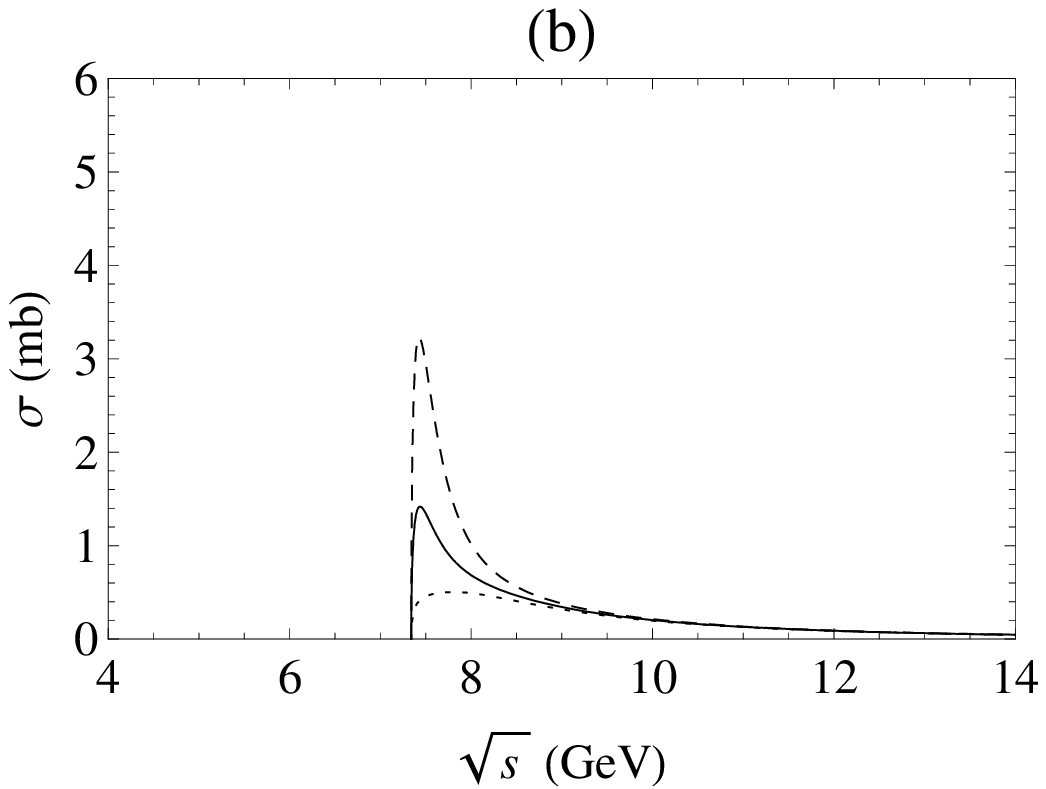} \label{fig4b}
\end{center}
\caption{$B_{c}$ absorption cross sections of the process $B_{c}^{+}\protect%
\pi \rightarrow D^{\ast }B^{\ast }$ for three different values of four-point
coupling, $g_{\protect\pi B_{c}B^{\ast }D^{\ast }}=105, 200, 295$ for dotted,
solid and dashed curve respectively (a) without and (b) with form factor.
Cutoff parameter is taken 1.5 GeV. }
\label{fig4}
\end{figure}

\bigskip

\noindent $B_{c}$ absorption cross section of the process $B_{c}^{+}\pi \rightarrow D^{\ast }B^{\ast }$
depends upon the four point contact coupling $g_{\pi B_{c}B^{\ast }D^{\ast }}$, whose values is fixed through SU(5)
symmetry. It is noted in the previous section that although SU(5) symmetry uniquely fix it, but difference in the values
of the couplings $g_{\pi DD^*}$ and $g_{\pi BB^*}$ produces two values 105 and 295 of the four point contact coupling.
In this paper, we treat this variation as uncertainty in the coupling and study its effect on the cross section of the process.
Fig. \ref{fig4}a, shows how the value of the four point
coupling could affect the values of $B_{c}$ absorption cross
sections through the process $B_{c}^{+}\pi \rightarrow D^{\ast }B^{\ast }$ without form factor.
Both of the cross sections increase very rapidly for the values 105 and 295,
which are not realistic. However, if we use the value of 200, the average to
two extreme values the variation in the cross section, denoted by solid line
is some what a compromise. Fig. \ref{fig4}b, shows the effect of uncertainty
in the four point contact coupling, on the cross section with form factor. This
figure indicates that the value of the contact coupling significantly affects
the cross section only near the threshold energy (7.34 GeV). It will be discussed later
that this effect is further marginalized in the total absorption cross section.

\begin{figure}[!h]
\begin{center}
\includegraphics[angle=0,width=0.70\textwidth]{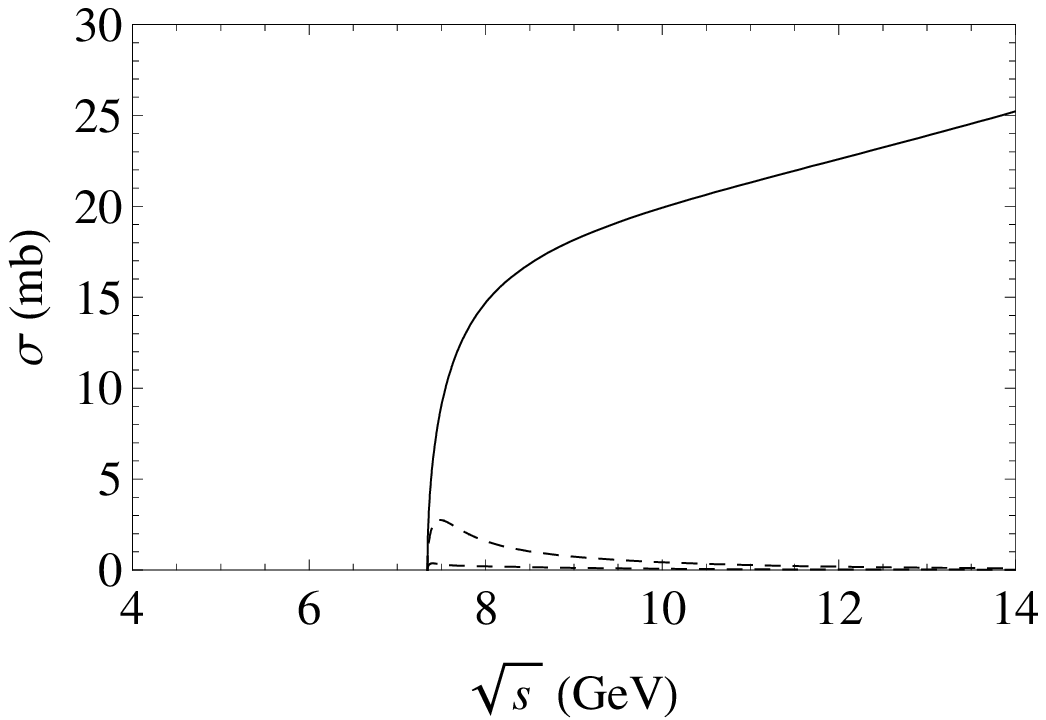}
\end{center}
\caption{$B_{c}$ absorption cross sections for the process $B_{c}^{+}\protect%
\pi \rightarrow D^{\ast }B^{\ast }$. Solid and dashed curves represents
cross sections without and with form factor respectively. Lower and upper
dashed curves are with cuttoff parameter $\Lambda =1$ and $2$ GeV
respectively and $g_{\pi B_{c}B^{\ast }D^{\ast }}=200$. Threshold energy is 7.34 GeV}
\label{fig5}
\end{figure}

\noindent Fig. 5 shows the $B_{c}$ absorption cross sections of the process $B_{c}^{+}\pi \rightarrow D^{\ast }B^{\ast }$
as a function of total center of mass (c.m) energy $\sqrt{s}$.
The cross section of the process roughly varies from 0.2 to 2 mb, when the cutoff
parameter is between 1 to 2 GeV and $g_{\pi B_{c}B^{\ast }D^{\ast }}=200$. Suppression due to form factor at cutoff \ $%
\Lambda =1$ and 2 GeV is roughly by factor 45 and 7 respectively. Relatively
high suppression in this process is mainly due to large values of mass of
final particles $D^{\ast }$ and $B^{\ast }$. It is noted that these
estimates of cross sections are highly dependent on the choice of form factor
and the value of cutoff, as well as on the values of coupling constants. However, it is observed that the effect of
uncertainty in the four point contact coupling $g_{\pi B_{c}B^{\ast }D^{\ast }}$ is marginal on the total cross section
due to relatively small value of the cross section of the second process. This is shown in the Fig. 6, in which
total absorption cross section for $B_{c}+\pi$ is plotted for three different values of $g_{\pi B_{c}B^{\ast }D^{\ast }}=105, 200, 295$.

\begin{figure}[!h]
\begin{center}
\includegraphics[angle=0,width=0.70\textwidth]{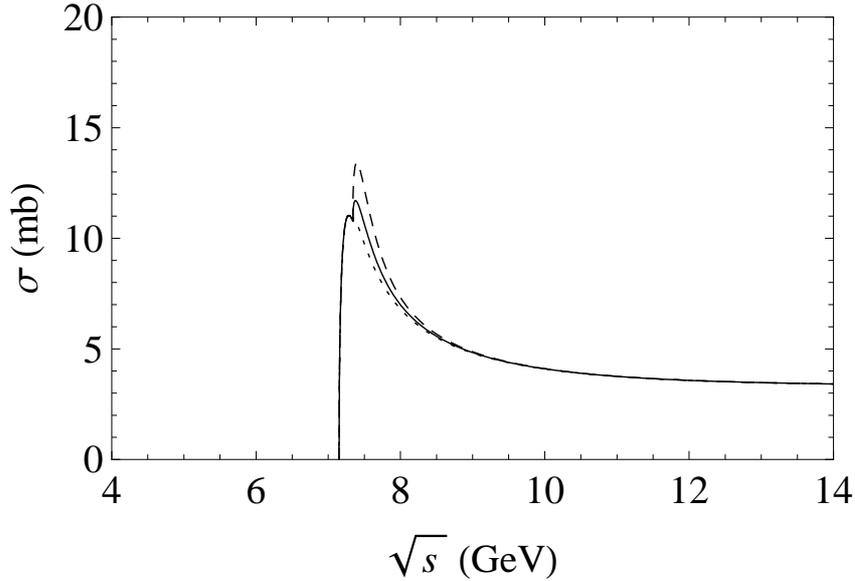}
\end{center}
\caption{Total $B_{c}$ absorption cross sections by pion for three different values of four-point
coupling, $g_{\protect\pi B_{c}B^{\ast }D^{\ast }}=105, 200, 295$ for dotted,
solid and dashed curve respectively. Cutoff parameter is taken 1.5 GeV.}
\label{fig6}
\end{figure}

\bigskip

\section{Concluding Remarks}

\noindent In this paper, we have calculated $B_{c}$ absorption cross section
by $\pi $ mesons using hadronic Lagrangian based on SU(5) flavor symmetry.
This approach has already been used for calculating absorption cross
sections of $J/\psi $ and $\Upsilon $ mesons by hadrons. In our study, all the coupling constants are preferably determined
empirically using vector meson dominance model, heavy quark symmetries or
QCD sum rules instead of using SU(5) symmetry. The hadronic Lagrangian based
on SU(5) flavor symmetry is developed by imposing the gauge symmetry,
but this symmetry is broken when the mass terms are added in the Lagrangian.
Thus SU(5) gauge symmetry exists only in limit of zero hadronic masses.
Broken SU(5) symmetry does not necessarily implies that the coupling
constants of three or four-point vertices should be related through SU(5)
universal coupling constant. It is, therefore, justified to empirically fix
the couplings. It can also be seen that the empirical values of the couplings
also violate SU(5) symmetry relations given in Eqs. 8 and 9. It is also
noted that four-point coupling constant $g_{\pi B_{c}B^{\ast }D^{\ast }}$
cannot be fixed empirically. Thus in this case we have no choice except to
make a reasonable estimate using SU(5) symmetry as discussed above. Calculated cross sections are found
to be in range 2 to 7 mb and 0.2 to 2 mb for the processes $B_{c}^{+}\pi
\rightarrow DB$ and $B_{c}^{+}\pi \rightarrow D^{\ast }B^{\ast }$
respectively, when the form factor is included. These results could be
useful in calculating production rate of $B_{c}$ meson in relativistic heavy ion
collisions.

\end{document}